\documentstyle[12pt]{article}

\begin{document}
{}~\hfill{UICHEP-TH/93-16}

{}~\hfill{10 Nov 1993}

\vspace{.6in}
\centerline {\large\bf Convergent WKB Series}
\vspace{.5in}

\centerline {David T. Barclay}
\vspace{0.3in}
\centerline {Department of Physics, University of Illinois at Chicago,}
\centerline {845 W. Taylor Street, Room 2236 SES,}
\centerline {Chicago, Illinois 60607-7059}
\vspace {1in}

\centerline{\bf Abstract}
\vspace{0.3in}

A set of simple exactly solvable potentials are shown to have convergent
WKB series.
The resulting all-orders quantisation conditions provide a
unified description of all known cases where an exact WKB
quantisation condition has been obtained by modifying
the potential ({\it \'a la} Langer), together with several
new examples.

\newpage

Investigations, both analytic [1] and numerical [2], into the higher-order
behaviour of the series (in $\hbar^2$) of corrections to the WKB
quantisation condition for particular potentials have usually shown the
series to be divergent.
This is in accord with the belief, on the basis of general properties
of differential equations, that the series is generically divergent,
though still asymptotic to the correct result [3].
On the other hand, there is the long-known fact that
the harmonic oscillator and Morse potentials have WKB series
with exemplary convergence: the lowest order quantisation condition
is exact and all the corrections are identically zero.
Additionally, Bender, Olaussen and Wang [2] have pointed out
that the Rosen-Morse potential
$V(x) = - V_0 / \cosh^2x$ has a series with a finite radius of convergence,
whose terms can be calculated and then summed to give a
simple all-orders quantisation condition.
Because the Rosen-Morse potential is an exactly solvable one, it can
be checked that the summed series gives the correct spectrum, i.e. that
no non-perturbative contributions such as terms like $e^{-1/\hbar}$
have been missed.
The main purpose of this paper is to show that this
convergent WKB series is not an isolated example
amongst simple exactly solvable potentials.

We approach the problem from a slightly different angle than Bender et
al by noting that the all-orders condition for Rosen-Morse could have
been deduced directly from the already known spectrum.
In general such a quantisation condition  will be of the form
$$ \int \sqrt{E-V} dx = (n + 1/2) \pi \hbar + \pi f(\hbar, E) \eqno(1) $$
where the function $f(\hbar,E)$ contains the $o(\hbar^2)$ and higher
corrections to the familiar WKB condition.
Provided one knows $E(n)$ exactly to begin with
and can evaluate the integral on the left hand side exactly, the only
unknown is $f$.
Deriving it this way automatically includes any non-perturbative terms.
Any possible query about the WKB series for $V(x)$ can be answered once
$f$ is known.

Unfortunately, while plenty of exactly solvable potentials are known,
$V(x)$ is usually sufficiently complicated that the integral is forbidding.
However it has proved possible to carry out this programme for a set
of simple, well-known potentials including the Coulomb, Eckart and
P\" oschl-Teller ones.
Introducing a function $u(x)$ such that $V= u^2$
(having fixed $V=0$ as the bottom of the well),
this set can be characterised as containing all the solutions to either
$$ {d u \over dx} = a +b u^2 +c u  \eqno(2)  $$
or
$$ {du \over dx} = a + b u^2 + cu \sqrt{a +b u^2},  \eqno(3) $$
where $a$, $b$ and $c$ are arbitary constants.
We should point out immediately that there is a close connection
here to the potentials discussed in [4] in connection with exact
lowest-order SWKB quantisation conditions (see below).
There equations (2) and (3) were satisfied by a superpotential $\phi$,
giving potentials
$$ V(x) = \phi^2 - \hbar \phi'+ \epsilon_0 = u^2, \eqno(4)  $$
where $\epsilon_0$ is the ground-state energy.
As detailed in the Appendix, any solution $u(x)$ to either (2) or (3)
with constants $a$, $b$ and $c$ corresponds to a potential $V(x)$ with
a $\phi(x)$ satisfying the {\it same} equation with some constants
$A$, $B$ and $C$.
Thus the set of potentials considered here is exactly the same set
discussed in [4]; while not essential to the argument, the fact that these
potentials have exact SWKB conditions greatly simplifies the algebra.
A minor exception aside [5], these potentials are usefully tabulated by
Dutt, Khare and Sukhatme [6].\footnote{These potentials all share the
elegant property of ``shape invariance'' (reviewed in [6]) and the emphasis
in [4] was on the relation between it and the SWKB series. Recent developments
[7] have weakened this relation, but do not effect the conclusions
in [4] concerning potentials for which the lowest-order SWKB condition
is exact.}

Note that equation (2) includes as solutions both the trivial harmonic
oscillator and Morse examples (for which $f=0$) and the Rosen-Morse
example of Bender et al.
This paper will thus directly generalise these earlier results.

We begin by proving perturbatively that for this set of potentials $f$ is
independent of $E$.
Of course this will become quite plain when we derive $f$ non-perturbatively,
but the proof does bring out some of the significance
of this set for the SWKB approximation.
Making the substitution $\psi= \exp(iS/\hbar)$, the Schr\"odinger equation
can be written as ($2m=1$)
$$ {S'}^2 - i \hbar S'' + V = E \eqno(5) $$
and the WKB approach involves expanding $S$ as a power series in $\hbar$,
the coefficients of which can be generated recursively [8].
Once connection formulae have been considered, one obtains a
quantisation condition which is a series in $\hbar^2$; in terms of $u(x)$
$$ \int \sqrt{E -u^2} dx - {\hbar^2 \over 6} {\partial^2 \over
  \partial E^2 } \int {u^2 {u'}^2 \over (E-u^2)^{1/2} } dx
    + \ldots = (n+1/2) \pi \hbar.  \eqno(6) $$
The higher-order corrections quickly increase in complexity, but any term
appearing in the $o(\hbar^{2n})$ one must have the structure
$$   {\partial^M \over \partial E^M } \int
   { u^{(n_1)} u^{(n_2)} \ldots u^{(n_m)} \over (E- u^2)^{1/2}}
				dx $$
$$ n_1 + n_2 + \ldots + n_m =2n,   \qquad M=(2n+m-2)/2  $$
$$   u^{(0)}=u, \quad u^{(1)} =u', \quad u^{(2)}=u'', \ldots \eqno(7) $$
It can also be arranged that $n_m=1$, so the integration variable
can be changed from $x$ to $u$.
Now notice that equations (2) and (3) have the unique property that
$$  {d^N u \over d x^N } = p(N+1) + p(N) g(u),  \eqno(8) $$
where $p(m)$ is used to denote an otherwise unspecified $m$th order
purely odd or even polynomial in $u$.
Solutions to (2) have $g(u)=1$ and those to (3) $g(u)=\sqrt{a+bu^2}$,
but no other choices of $g$ repeat this pattern (8).
For these cases, (7) becomes
$$ {\partial^M \over \partial E^M}
   \int_{-\sqrt E}^{\sqrt E} {p(2M) + p(2M-1) g(u)
   \over (E-u^2)^{1/2} } du. \eqno(9) $$
The term containing $g(u)$ is odd and integrates to zero; the other is a
polynomial in $u^2$ with leading term $u^{2M}$, so that the integral gives
a polynomial in $E$ with leading term $E^M$.
Differentiating $M$ times leaves a constant independent of $E$.
Since this happens in  all terms in the series, for these potentials
$f(\hbar,E)$ reduces to $f(\hbar)$ with no $E$-dependence.
Their all-orders quantisation condition (1) is thus particularly similar
to the standard lowest-order one: the $E$-dependence is exactly the same
and the higher-orders can be treated as an additional constant on the
right hand side, absorbable into a shifted $n$.\footnote{Potentials with
$f$ independent of $E$ are the only ones for which the semi-classical
inverse method [9] is exact. This is because differentiating (1) with
respect to $E$ (the first step therein) then gives a formula indistinguishable
from the lowest-order one. Given $n(E)$, one can thus reconstruct the
``excursion'' of $V(x)$ exactly.}
There may be other potentials with $E$-independent higher-orders, but
experience with the SWKB series suggests otherwise [4].

Note that the SWKB series would be derived by splitting $V$ into $\phi^2
-\hbar \phi'$ and treating the second piece as $o(\hbar)$ [10].
This leads to a slightly different approximation scheme with a series
somewhat like (6), but with integrals over $\phi$ rather than $u$.
The arguments above go through rather similarly, except that the
presence of a factor of $E-\epsilon_0$ multipying each correction (the SWKB
approximation is exact as $E \rightarrow \epsilon_0$) means that the final
integral
is over a polynomial $p(2M-2)$ and thus that the $M$ derivatives leave
nothing.
Each correction is zero and these potentials have exact lowest-order
SWKB conditions [4]
$$ \int \sqrt{E-\epsilon_0- \phi^2} dx = n \pi \hbar. \eqno(10) $$

It is this fact that allows for a particularly simple derivation of $f$
for these potentials, avoiding the necessity of calculating it separately
for each individual potential.
Because of (10), we can rewrite (1) as
$$ \int {\sqrt{E- u^2} \over u'} du =
   \int {\sqrt{E-\epsilon_0-\phi^2} \over \phi'} d\phi + { \pi \hbar \over 2} +
\pi f.
            \eqno(11) $$
But we know $u'$ in terms of $u$ from (2) and (3) and similarly for $\phi'$,
so the integrals are seen to be elementary ones.
That on the right gives a function of $A$, $B$ and $C$, but this can be recast
in terms of $a$, $b$ and $c$ using the formulae in the Appendix.
For potentials corresponding to (2) the result is
$$ f = \sqrt{ {1 \over b^2} + {\hbar^2 \over 4} } - {1 \over b}. \eqno(12) $$
As expected, the $E$-dependence has cancelled and $f \rightarrow 0$ as
$b \rightarrow 0$, the latter to give agreement with the known absence
of higher-order corrections for the harmonic oscillator and Morse potentials.
Setting $b=1/\sqrt{V_0}$ shows (12) to be in agreement with the Bender et al
result [2].

The equivalent function for potentials given by (3) is more complicated.
$$ f = {1 \over b-c^2} \bigl( H(\hbar^2) -1 \bigr)  $$
$$ H^2(\hbar^2) \equiv {1 \over 8} \Bigl( 4 b^{-1} (b+c^2) +
  \hbar^2 (c^2-b)^2 - (c^2-b) \sqrt{(4/b + \hbar^2 b + \hbar^2 c^2)^2
    -4 \hbar^4 c^2 b} \Bigr).  \eqno(13) $$
Note that $H(\hbar^2) \rightarrow 1$ as $c^2 \rightarrow b$, so that, contrary
to appearances, the $c^2=b$ case is not pathological; this corresponds to
the radial equation for the 3D harmonic oscillator and has
$$ f = {1 \over 4b} \bigl( \sqrt{ 1 + \hbar^2 b^2} -1 \bigr).  \eqno(14) $$

These last equations, (12)-(14), are the central results of this paper.
All these functions are analytic about $\hbar=0$.

Equation (12) is particularly important in that it sheds new light on a
classic body of work concerned with modifying potentials such that the lowest
order WKB condition for this new potential gives the correct spectrum for
the unmodified one.
This can clearly be done in principle for any potential, but a
surprisingly simple modification was discovered for the radial version
of the Coulomb potential very early [11]: replace $l(l+1)$ in the
centrifugal barrier by $(l+1/2)^2$.
Langer [12] subsequently showed that a consistent treatment of the WKB
approximation for radial problems introduces an additional term in
the lowest order condition that is just equivalent to this modification.
But this is a demonstration that the condition for which the
Coulomb potential is exact is the correct lowest order one rather than
an explanation for this exactness.
The first proof (excluding empirical observation that the trick works) was
by Rosenzweig and Krieger [13] within the Froman version [14] of the
approximation.
However this was only as one of five case-by-case examples
of potentials which could be modified to give the correct answer, some
of which were unrelated to Langer's work, e.g. the Rosen-Morse
potential $V(x)=-V_0/\cosh^2x$ can be modified using [15]
$$  {V'}_0 = V_0 + {\hbar^2 \over 4}, \eqno(15) $$
but is an innocuous 1D problem.

Surprisingly, all the potentials considered by Rosenzweig and Krieger are
members of the set considered above and so it is possible to bring together
their stray examples into one generalisation.
Before turning to this, we should however deal with the possible objection
that Langer's observations invalidate the derivations above for particular
cases, notably the Coulomb one. For instance, the all-orders result
appears to imply a lowest-order quantisation condition in disagreement
with Langer's for the Coulomb problem.
Actually, because Langer's modified potential is invariably $\hbar$-dependent,
there need not necessarily be any contradiction between the two lowest-order
formulae; the modified condition may result from a partial summation of
higher-order pieces in the original series (6).
And for the set of potentials here that series (6) can always be rendered
meaningful by appealing to their exact SWKB conditions, since
the WKB series (6) can always be obtained by rearranging the SWKB
one (c.f. [16]).
Thus na\"\i vely applying (6) to the Coulomb potential gives a series
in $\hbar$ which will sum to the correct answer and uniqueness requires
that this be the same series implied in (12).
Of course no such justification is necessary for potentials like the
Rosen-Morse one which have no singularities.

Rewrite (2) as
$$ u' = b(u+p)(u+q).  \eqno(16) $$
Hitherto the arbitrary reference energy has been chosen such that $V=0$
at the minimum of the potential, but this minimum is one
of the things that can be expected to be modified,
so we introduce $\tilde E \equiv E-p^2$.
For the Coulomb case this corresponds to refering all energies to the top
of the well.
{}From (1) and (12), the exact condition is
$$ {-1 \over b} + {1 \over b(p-q)} \Bigl[ q\sqrt{1 - (\tilde E+p^2)/q^2}
   -p \sqrt{ 1 -(\tilde E+ p^2)/p^2} \Bigr] = (n+1/2) \pi \hbar
       + \sqrt{ {1 \over b^2 } + {\hbar^2 \over 4}}  \eqno(17) $$
and we seek modified $b'$, $p'$ and $q'$ such that
$$ {-1 \over b'} + {1 \over b'(p'-q')} \Bigl[ q'\sqrt{1 -
    (\tilde E+{p'}^2)/{q'}^2}
   -p' \sqrt{ 1 -(\tilde E+ {p'}^2)/{p'}^2} \Bigr] = (n+1/2) \pi \hbar.
        \eqno(18) $$
Thus
$$ {1 \over {b'}^2} = {1 \over b^2} + {\hbar^2 \over 4}$$
$$ q^2 -p^2 = {q'}^2 - {p'}^2, \qquad  b(p-q) = b' (p'-q'). \eqno(19) $$
In the Coulomb case
$$ b={1 \over \hbar \sqrt{l(l+1)}}, \qquad p=q={-e^2 \over 2 \hbar
                                          \sqrt{l(l+1)}} \eqno(20) $$
and, as expected, (19) implies that the substitution
$l(l+1) \rightarrow (l+1/2)^2$ gives $b'$, $p'$ and $q'$.
As a example of a new result obtained, consider
$$ V(x) = {-V_0 \over \cosh^2x } + 2 \sqrt{V_0} \delta \tanh x + \delta^2,
     \eqno(21) $$
a more general version of the Rosen-Morse potential dealt with by
Rosenzweig and Krieger [13] and Bender et al [2].
The modified potential is given by (15), now supplemented with
$$ \delta' = \sqrt{ {V_0 \over {V'}_0 }} \delta. \eqno(22)  $$
All the modifications tabulated in [14] are subsumed into (19), along with
several such generalisations.

To summarise, we have shown that potentials $V=u^2$ given by solutions
to (2) and (3) have -- in addition to exact lowest-order SWKB conditions
(10) -- WKB series (6) which have finite radii of convergence and which
sum to give the particularly simple quantisation conditions corresponding
to (12) and (13) respectively.
For the potentials with singularities, Langer's modification of the
lowest-order WKB formula has been seen to be equivalent to the summation
of the higher-order corrections in the original (and here ostensibly incorrect)
version of the approximation.
In addition, equations (19) provide a recipe for modifying the potentials
-- whether containing singularities or not -- in all those cases where such
simple modifications have given the correct spectrum in lowest-order WKB,
together with some new examples.

This sample of potentials is obviously very special
and the immediate challenge is clearly to find $f(\hbar,E)$ for other
exactly solvable potentials.

\vspace{0.3in}

{\large\bf Appendix}

\vspace{0.3in}

Consider two functions $u(x)$ and $\phi(x)$ related by
$$ \phi^2 - \hbar \phi'+\epsilon_0 = u^2.  \eqno(A1) $$
If $\phi$ is a solution to
$$ \phi' = A + B \phi^2  + C \phi,   \eqno(A2) $$
then $u$ is a solution to (2) with
$$ a= (1- \hbar B)^{1/2}A + {\hbar C^2 \over 4} {(2-\hbar b) \over
		(1-\hbar B)^{3/2}} $$
$$ b=B(1-\hbar B)^{-1/2}, \qquad c=C(1-\hbar B)^{-1} $$
$$ \epsilon_0 = {\hbar^2 C^2 \over 4(1-\hbar B)} + \hbar A. \eqno(A3) $$
Similarly, if $\phi$ is a solution to
$$ \phi' = A + B\phi^2 + C \phi \sqrt{A + B \phi^2},  \eqno(A4)$$
then $u$ is a solution to (3) with
$$ a= \alpha \sqrt{AB} - \beta C \sqrt{A}, \qquad b=a(\alpha^2 -
   \beta^2)^{-1} $$
$$ c= b^{-1/2} ( \beta \sqrt{AB} - \alpha C \sqrt{A}) $$
$$ \epsilon_0 = \hbar A + {A(1 -\hbar B) \over 2B} \Bigl( 1 -
		\sqrt{1- \hbar^2 C^2 B / (1- \hbar B)^2 } \Bigr),  \eqno(A5)$$
where $\alpha^2 \equiv A/B - \epsilon_0$ and $\beta^2 \equiv \epsilon_0
- \hbar A$.

In both cases the appropriate converse is also true.

\vspace{0.3in}

{\large\bf Acknowledgments}

\vspace{0.3in}

I would like to thank C.J. Maxwell and A.V. Turbiner for helpful
comments.

\vfill
\eject

\end{document}